\documentclass{appolb}
\usepackage{epsfig}
\usepackage{color}
\usepackage{cite}
\usepackage{color}
\usepackage{amsfonts}

\begin{document}
\title{Anomalous diffusion models: different types of subordinator distribution
}
\author{Joanna Janczura and Agnieszka Wy{\l}oma{\'n}ska
\address{Hugo Steinhaus Center,
    Institute of Mathematics and Computer Science,\\
    Wroclaw University of Technology, Poland}
}
\maketitle
\begin{abstract}
Subordinated processes play an important role in modeling anomalous diffusion-type behavior. In such models the observed constant time periods are described by the subordinator distribution. Therefore, on the basis of the observed time series, it is possible to conclude on the main properties of the subordinator. In this paper we analyze the anomalous diffusion models with three types of subordinator distribution: $\alpha-$stable, tempered stable and gamma. We present similarities and differences between the analyzed processes and point at their main properties (like the behavior of moments or the mean squared displacement).
\end{abstract}
  
\section{Introduction}
Brownian motion is a classical continuous-time model describing diffusion of particles in some fluid. Besides physics, it has found many real-world applications, like in ecology, medicine, finance and many other fields \cite{ein1}.
 
But in spite of many obvious advantages, the standard Brownian diffusion cannot model the real time series with apparent constant time periods (called also trapping events), which are often observed in datasets recorded within various fields. Therefore a rapid evolution of alternative models is observable in many areas of interest. Especially anomalous diffusion models have found many practical applications. They were used in variety of physical systems, including charge carrier transport in amorphous semiconductors \cite{SM,SL,PS}, transport in micelles \cite{OBLU}, intracellular transport \cite{CGE} or motion of mRNA molecules inside E. coli cells \cite{GC}. The constant time periods can be also observed in processes corresponding to stock prices or interest rates, so models based on the subordinated processes might be also useful in modeling financial time series, \cite{janczura_wylo, janczura_orzel_wylo, orzel_wylo}. 

One of the most important issues that arises in the analysis of the subordinated processes is the description of waiting-times that correspond  to the periods of constant values. Finding a proper subordinator distribution allows to conclude on the properties of the whole process. The most popular subordinator distribution is the inverse $\alpha-$stable, see for instance \cite{mag:wer:07}, but recent developments in this area indicate that another nonnegative infinitely divisible distribution can be also used to model the observed waiting-times, \cite{orzel_wylo,gaj:mag:10,mag1}.  The family of such distributions contains, besides one-sided L\'evy stable, also Pareto, gamma, Mittag-Leffler or tempered stable.

In this paper we analyze the subordinated Brownian motion with three types of the inverse subordinator distribution, namely $\alpha-$stable, tempered stable and gamma. We show the differences between the distributions and present the main properties of the analyzed subordinated processes mainly expressed in the language of moments. Moreover, we investigate the asymptotics of the mean squared displacement and show that in the gamma case it is linear for large $t$, while for small $t$ it exhibits non-power law behavior. 

\section{Subordinated Brownian motion}
We start with introducing a general definition of the considered processes. The subordinated Brownian motion is defined as \cite{mag1}:
\begin{eqnarray}\label{sub1}
Y(t)=B(S(t)),
\end{eqnarray}
where $\{B(\tau)\}_{\tau\geq 0}$ is the Brownian motion and $\{S(t)\}_{t\geq 0}$ is an inverse subordinator of $\{U(\tau)\}_{\tau\geq 0}$ \cite{woy,mag5}, i.e.:
\begin{eqnarray}\label{salpha}
S(t)=\inf\{\tau>0: U(\tau)>t\}\end{eqnarray}
for  increasing L\'evy process $\{U(\tau)\}_{\tau\geq 0}$ with the Laplace transform given by:
\begin{eqnarray}\label{psi}
E e^{-zU(t)}=e^{-t\Psi(z)}.
\end{eqnarray}
The function $\Psi(z)$ is called the L\'evy exponent and can be written in the following form
\[\Psi(z)=\theta z+\int_{0}^{\infty}(1-e^{-zx})v(dx).\]
Here, $\theta\geq 0$ is the drift parameter. If for simplicity, following \cite{mag1}, we assume $\theta=0$, then  $v(dx)$ is an appropriate L\'evy measure.
Moreover, $B(\tau)$ and $S(t)$ are assumed to be independent.  

The probability density function (pdf) of the process $\{Y(t)\}_{t\geq 0}$ is characterized by the generalized Fokker-Planck equation \cite{mag1,sok}:
\begin{eqnarray}\label{FPE}
\frac{\partial w(x,t)}{\partial t}=\frac{1}{2}\frac{\partial^2}{\partial x^2}\Phi  
w(x,t),~~w(x,0)=\delta(x),
\end{eqnarray}
where $\delta(x)$ is the Dirac delta in point $x$ and $\Phi$ - an integro-differential operator defined as:
\begin{eqnarray}\label{phi}
\Phi f(t)=\frac{d}{dt}\int_{0}^{t}M(t-y)f(y)dy.
\end{eqnarray}
The function $M(t)$ is called the memory kernel and is defined via its Laplace transform, \cite{mag1}:
\[\tilde{M}(z)=\int_{0}^{\infty}e^{-zt}M(t)dt=\frac{1}{\Psi(z)}.\]

\section{Three cases of subordinator distribution}
The classical anomalous diffusion type model given by the subordinated Brownian motion (\ref{sub1}) defines subordinator $S(t)$ as an inverse $\alpha$-stable process, see for instance \cite{mag5}. It implies that the lengths of the constant time periods are $\alpha$-stable distributed. However, in some applications also different distributions describing the lengths of the constant time periods might be useful. In this paper, besides $\alpha$-stable, we consider two other cases of subordinator distribution, namely tempered stable $TS(\alpha,\lambda,c)$ and gamma $G(a,c)$. 

We start with a brief review of the main properties of the considered distributions.

\subsection*{$\alpha$-stable distribution}

Since there is no closed form for the probability density function of the $\alpha$-stable distribution, it is usually more conveniently defined by it's characteristic function, given by 
\begin{equation}
\psi(t)=\left\{ \begin{array}{lcr} \exp\left\{-\sigma^{\alpha}|t|^{\alpha}\left[ 1-i\beta \mbox{sign}(t)\tan \frac{\pi\alpha}{2} \right]+i\mu t \right\} & \mbox{ if } \alpha\neq 1,\\
\exp\left\{-\sigma|t|\left[ 1+i\beta \mbox{sign}(t) \frac{2}{\pi} \ln|t| \right]+i\mu t \right\} & \mbox{ if } \alpha=1, \end{array}
\right.
\end{equation}
where $\alpha\in(0,2]$ is the stability parameter, $\beta\in[-1,1]$ is the skewness parameter, $\sigma>0$ is the scale parameter and $\mu \in \mathbb{R}$ is the location parameter. Note that if $\alpha<1$ and $\beta=1$ the stable distribution becomes totally (right) skewed. Since the subordinator should be a non-decreasing process, in the following we assume $\alpha<1$, $\beta=1$ and $\mu=0$. Moreover, for simplicity we assume $\sigma=1$.

Recall, that the $\alpha-$stable family has two important properties. First, a sum of two independent $\alpha$-stable random variables with the same $\alpha$ parameter is again $\alpha$-stable distributed. Second, the tails of the stable distribution are governed by the power law behavior. 
 
\subsection*{Tempered stable distribution}
The positive tempered stable random variable $T$ with parameters $\alpha,\lambda$ and $c$ is defined through the Laplace transform
\begin{equation}
E\left(e^{-uT}\right)=e^{-c((\lambda+u)^{\alpha}-\lambda^{\alpha})},~ \lambda>0,~ 0<\alpha<1,~c>0.
\end{equation}
In the above definition $\lambda$ is the tempering parameter, while $\alpha$ and $c$ are the stability and scale parameters, respectively. Observe that if $\lambda=0$, then the random variable $T$ becomes simply $\alpha-$stable with the scale parameter $c^{1/\alpha}$. The  probability density function (pdf)  $p^{TS}_{\alpha,\lambda,c}$ of the tempered stable distribution with parameters $\alpha,\lambda$ and $c$ can be expressed in the following form:
\begin{eqnarray}\label{pdf2}
p^{TS(\alpha,\lambda,c)}(x)=e^{-\lambda x+c\lambda^{\alpha}}p^{S(\alpha,\sigma,1,0)}(x),
\end{eqnarray}
where $\sigma=\left(c*\cos\frac{\pi\alpha}{2}\right)^{1/\alpha}$ and $p^{S(\alpha,\sigma,\beta, \mu)}(x)$ is the pdf of the $\alpha$-stable distribution with the stability index $\alpha$, scale parameter $\sigma$, skewness $\beta$ and shift $\mu$, \cite{maersh,orzel_wylo,gaj:mag:10}. Because
all moments of the tempered stable distribution are finite, it becomes attractive in many practical applications for instance in finance \cite{kim,kim1}, biology \cite{hou} and physics like anomalous diffusion, relaxation phenomena \cite{stanis,gaj:mag:10}, turbulence \cite{chec1} or plasma physics \cite{chec2b}, see also \cite{chec3,chec4}.  

\subsection*{Gamma distribution}

The pdf of the gamma distribution $p^{G(a,c)}$ is given by
\begin{equation}
p^{G(a,c)}(x)=x^{c-1}\frac{e^{-x/a}}{\Gamma(c)a^{c}},~x\geq0,
\end{equation}
where $\Gamma(z)$ is the Gamma function defined as:
\begin{equation}
\Gamma(z)=\int_{0}^{\infty} t^{z-1}e^{-t}dt.
\end{equation}
It is interesting to note that for $a=1$ the gamma distribution becomes the exponential one. Moreover, gamma distribution is infinitely divisible. For $X_i\sim G(a_i,c)$ we have $\sum_{i=1}^{n} X_i \sim G(\sum_{i=1}^n a_i,c)$ provided that $X_i$ are independent.

\begin{figure}[tb]
\begin{center}
\includegraphics[width=12cm]{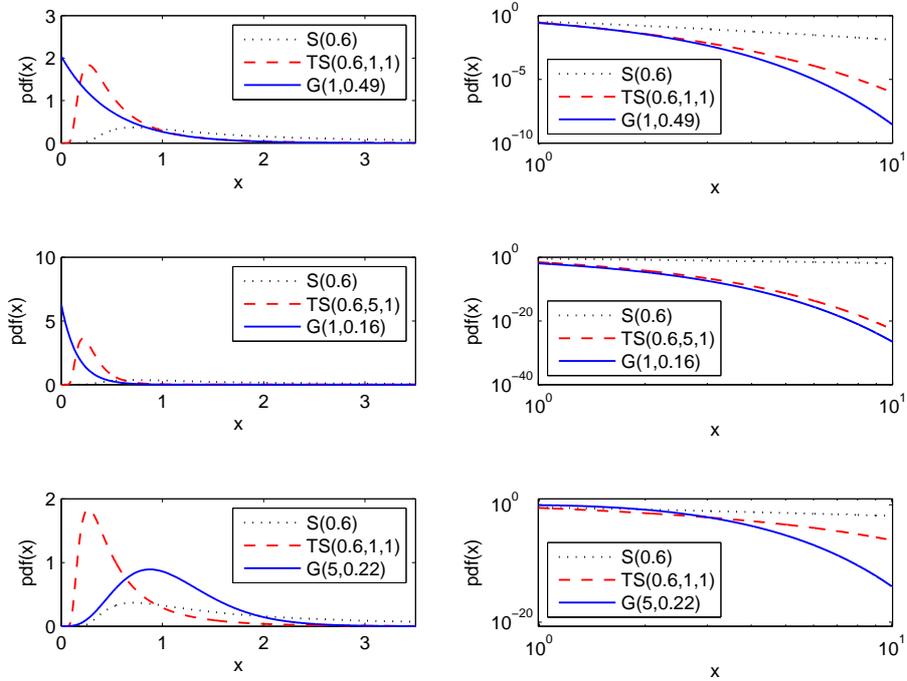}\label{pdf}
 \caption{Probability density functions (pdf) of the three considered distributions: S($\alpha$) -- $\alpha$-stable, TS($\alpha,\lambda,c$) -- tempered stable and G(a,c) -- gamma. The $\alpha$ parameters in the stable and tempered stable distributions are equal to $0.6$. The parameters of the gamma distribution are chosen so it's mean is equal to the mean of the tempered stable distribution. Right panels display the right tails of the corresponding probability density functions in the double-logarithmic scale.}
\end{center}
\end{figure}

In Figure \ref{pdf} we plot sample probability density functions, as well as, the tails of the considered distributions. The $\alpha$ parameters in stable and tempered stable distributions are equal to $0.6$. The parameters of the gamma distribution are chosen so it's mean is equal to the mean of the tempered stable distribution.

\section{Subordinated Brownian motion with different types of subordinator distributions}
In this section we examine the subordinated Brownian motion defined in (\ref{sub1}) with three types of subordinator distribution, namely: $\alpha-$stable, tempered stable and gamma.

\subsection*{$\alpha-$stable case}
The $\alpha$-stable subordinator $\{U_{\alpha}(\tau)\}_{\tau\geq 0}$ is a non-decreasing L\'evy process with the L\'evy measure $v(dx)=x^{-(1+\alpha)}I_{x>0}dx$ and the following Laplace transform:
\begin{eqnarray}\label{tr_stable}
Ee^{-zU(t)}=e^{-t z^\alpha},~~ 0<\alpha<1.
\end{eqnarray}
Therefore the function $\Psi$, that appears in (\ref{psi}), takes the form
\begin{equation}
\Psi(z)=z^{\alpha}.
\end{equation}
This implicates the form of the memory kernel, namely
\begin{equation}
M(t)=\frac{t^{\alpha-1}}{\Gamma(\alpha)}.
\end{equation}
The first two moments of the subordinated Brownian motion defined in (\ref{sub1}) in the $\alpha$-stable case are given by \cite{janczura_wylo}:
\begin{eqnarray}\label{mom1}
<Y(t)>=0,~~<Y^2(t)>=\frac{1}{\Gamma(\alpha+1)}t^{\alpha},
\end{eqnarray}
while the covariance function takes the form
\begin{eqnarray}\label{cov}
<Y(t),Y(s)>=\frac{\min(s,t)^{\alpha}}{\Gamma(\alpha+1)}.
\end{eqnarray}

\subsection*{Tempered stable case}
The tempered-stable subordinator  $\{U_{\alpha,\lambda,c}(\tau)\}_{\tau\geq 0}$ is a L\'{e}vy process with tempered stable increments (i.e.\ with the L\'evy measure $v(dx)=\frac{e^{-\lambda x}}{x^{1+\alpha}}I_{x>0}dx$) and the Laplace transform given by: \cite{stanis}
\begin{eqnarray}\label{laptr}
E\left(e^{-zU(t)}\right)=e^{-t\Psi(z)}=e^{-t((\lambda+z)^{\alpha}-\lambda^{\alpha})},
\end{eqnarray}
where $\lambda>0$, $0<\alpha<1$. Let us point out that in case $\lambda\rightarrow0$ the  operator $\Phi$  is proportional to the fractional Riemann-Liouville derivative,
therefore (\ref{pdf}) tends to fractional Fokker-Planck equation (see subsection $\alpha$-stable case), \cite{stanis}.
The basic properties and the simulation procedure of the process $\{Y(t)\}$ defined in (\ref{sub1}) in the tempered stable case one can find in \cite{mag1,gaj:mag:10,orzel_wylo,woy}.

Observe that the memory kernel $M(t)$ in the considered case can be calculated on the basis of the following equation
\begin{eqnarray}\label{memory}
\int_{0}^{\infty}e^{-ut}M(t)dt=\frac{1}{\Psi(u)}=\frac{1}{(\lambda+u)^{\alpha}-\lambda^{\alpha}}.
\end{eqnarray}
As a consequence, the memory kernel $M(t)$ takes the form:
\begin{eqnarray}\label{memory1}M(t)=e^{-\lambda t}t^{\alpha-1}E_{\alpha,\alpha}((\lambda t)^{\alpha}),\end{eqnarray}
where
\[E_{\alpha,\beta}(z)=\sum_{k=0}^{\infty}\frac{z^k}{\Gamma(\alpha k+\beta)}\] is a generalized Mittag-Leffler function, \cite{golou}. 

Since for $0<\alpha<1$ and $\beta<1+\alpha$, the generalized Mittag-Leffler function for $z\in R $ and $z\neq 0$ can be expressed as (see Theorem 2.3 in \cite{golou}):
\[E_{\alpha,\beta}(z)=\frac{1}{\alpha}z^{(1-\beta)/\alpha}\exp\{z^{1/\alpha}\}+\int_{0}^{\infty}K(\alpha,\beta,r,z)dr,\]
where
\begin{equation}
K(\alpha,\beta,r,z)=\frac{1}{\pi\alpha}r^{(1-\beta)/\alpha}e^{-r^{1/\alpha}} \frac{r\sin(\pi(1-\beta))-z\sin(\pi(1+\beta-\alpha))}{r^2-2rz\cos(\pi\alpha)+z^2},
\end{equation}
the memory kernel $M(t)$ is given by the following formula:
\begin{eqnarray}\label{Mt2}
M(t)=\frac{\lambda^{1-\alpha}}{\alpha}+e^{-\lambda t}t^{\alpha-1}\int_{0}^{\infty}K(\alpha,\alpha,r,(\lambda t)^{\alpha})dr.
\end{eqnarray}
Therefore we have
\begin{equation}
\lim\limits_{t\rightarrow\infty}M(t)=\frac{\lambda^{1-\alpha}}{\alpha}.
\end{equation}
The above limiting behavior is a simple consequence of the fact that for large $t$ the generalized Mittag-Leffler function can be written as \cite{golou}:
\begin{equation}
E_{\alpha,\alpha}((\lambda t)^{\alpha})=\frac{(\lambda t)^{1-\alpha}e^{\lambda t}}{\alpha}-\sum_{k=0}^p\frac{(\lambda t)^{-k}}{\Gamma(\alpha(1-k))}+O(|\lambda t|^{-1-p}).
\end{equation}


Knowing the form of the memory kernel we can calculate the basic statistics of the process $\{Y(t)\}$ such as moments and autocovariance function (see Theorem 1 in \cite{mag1}), namely
\begin{equation}
<Y(t)>=0,~~<Y^2(t)>=\int_{0}^tM(u)du
\end{equation}
and
\begin{equation}
<Y(t),Y(s)>=\int_{0}^{\min(t,s)}M(u)du.
\end{equation}
However, in case of the tempered stable distribution such derivations require numerical approximations.

\subsection*{Gamma case}
The gamma subordinator $\{U_{a,c}(\tau)\}_{\tau\geq 0}$ is a L\'{e}vy process with independent gamma distributed increments, i.e.\ with the L\'evy measure $v(dx)=c\frac{e^{-a x}}{x}I_{x>0}dx$) and the Laplace transform given by:
\begin{eqnarray}\label{laptr_gama}
E\left(e^{-zU(t)}\right)=\left(\frac{1}{1+za}\right)^{ct},~~a>0,~~c>0.
\end{eqnarray}
Observe that in this case also the one-dimensional density $p(t,x)$ of the process $U_{a,c}(\tau)$ is given in a closed form, namely
\begin{equation}\label{eqn:gamma:pdf}
p(t,x)=x^{ct-1}\frac{e^{-x/a}}{\Gamma(ct)a^{ct}}.
\end{equation}
In this case the L\'evy exponent $\Psi(z)$ is given by:
\begin{equation}
\Psi(z)=c\log(1+za),
\end{equation}
what implicates that the memory kernel $M(t)$ can be expressed as:
\begin{equation}
M(t)=\textit{L}^{-1}\left(\frac{1}{c\log(1+za)}\right),
\end{equation}
where $\textit{L}^{-1}(f(t))$ is the inverse Laplace transform of the $f(t)$ function. In order to find a formula for the memory kernel $M(t)$, we use the following relation (being a consequence of the Proposition 1 in \cite{mag1}):
\begin{equation}\label{eqn:var:m}
<Y(t)^2>=\int_0^{t}M(u)du.
\end{equation}
On the other hand, we have
\begin{equation}\label{eqn:Y2:S}
<Y(t)^2>=<S(t)>,
\end{equation}
where $\{S(t)\}$ is the inverse subordinator. Moreover, using the relation between subordinator $\{U(\tau)\}$ and its inverse and the fact that for each $t$ the random variable $S(t)$ is positive, we obtain
\begin{equation}
<S(t)>=\int_0^{\infty}P(S(t)>\tau)d\tau=\int_0^{\infty}P(U(\tau)\leq t)d\tau.
\end{equation}
Therefore, in case of the gamma distribution, we get
\begin{equation}\label{eqn:S}
<S(t)>=\int_0^{\infty}\frac{\gamma(c\tau,t/a)}{\Gamma(c\tau)}d\tau,
\end{equation}
where $\gamma(s,x)$ is an incomplete gamma function defined as:
\begin{equation}
\gamma(s,x)=\int_{0}^x t^{s-1}e^{-t}dt.
\end{equation}
Finally, from (\ref{eqn:var:m}) we have
\begin{equation}
M(t)=\frac{1}{c}\int_0^{\infty}\frac{1}{\Gamma(\tau)}\frac{\partial \gamma(\tau,t/a)}{\partial t}d\tau=\frac{e^{-t/a}}{c}\int_0^{\infty}\frac{t^{\tau-1}}{a^{\tau}\Gamma(\tau)}d\tau.
\end{equation}

Again, the basic statistics of the process $Y(t)$ can be calculated. Observe that from (\ref{eqn:Y2:S}) and (\ref{eqn:S}) we have
\begin{equation}
<Y(t)^2>=\int_{0}^{\infty}\frac{\gamma(c\tau,t/a)}{\Gamma(c\tau)}d\tau
\end{equation}
and
\begin{eqnarray}
<Y(t),Y(s)>&=&<B^2(\min(s,t)>=<S(\min(s,t)>=\\
&=&\int_{0}^{\infty}\frac{\gamma(c\tau,\min(s,t)/a)}{\Gamma(c\tau)}d\tau.
\end{eqnarray}

The main characteristics of the subordinated process $\{Y(t)\}_{t\geq 0}$ defined in (\ref{sub1}) for the three considered cases of subordinator distribution are summarized in Table \ref{tab:char}.

\begin{table}
\begin{center}
\scriptsize
 \caption{Characteristics of the subordinated process $\{Y(t)\}_{t\geq 0}$ defined in (\ref{sub1}) for the three cases of the subordinator distribution.}\label{tab:char}\vspace*{0.5cm}
\begin{tabular}{|c|c|c|c| }
\hline
Subordinator  &    Stable &     Tempered stable &     Gamma \\
distribution & S($\alpha$) & TS($\alpha,\lambda,c$) & G($a,c$)\\\hline
Parameters&$\alpha$&$\alpha,\lambda,c$&$a,c$\\\hline
$\Psi(z)$&$z^{\alpha}$&$c\left((\lambda+z)^{\alpha}-\lambda^{\alpha}\right)$&$c\log(1+az)$\\
&&&\\\hline
$M(t)$&$\frac{t^{\alpha-1}}{\Gamma(\alpha)}$&$\frac{1}{c}e^{-\lambda t}t^{\alpha-1}E_{\alpha,\alpha}((\lambda t)^{\alpha})$&$\frac{e^{-t/a}}{c}\int_{0}^{\infty}\frac{t^{\tau-1}}{a^{\tau}\Gamma(\tau)}d\tau$\\
&&&\\\hline
$<Y(t)^2>$&$\frac{t^{\alpha}}{\Gamma(\alpha+1)}$&$\frac{1}{c}\int_{0}^te^{-\lambda u}u^{\alpha-1}E_{\alpha,\alpha}((\lambda u)^{\alpha})du$&$\int_{0}^{\infty}\frac{\gamma(c\tau,t/a)}{\Gamma(c\tau)}d\tau$\\
&&&\\\hline
$<Y(t),Y(s)>$&$\frac{\min(t,s)^{\alpha}}{\Gamma(\alpha+1)}$&$\frac{1}{c}\int_{0}^{\min(t,s)}e^{-\lambda u}u^{\alpha-1}E_{\alpha,\alpha}((\lambda u)^{\alpha})du$&$\int_{0}^{\infty}\frac{\gamma(c\tau,\min(s,t)/a)}{\Gamma(c\tau)}d\tau$\\
&&&\\\hline
\end{tabular}
\end{center}
\end{table}



\section{Sample paths properties of the subordinated Brownian motion with different types of subordinator distribution}

\begin{figure}[tb]
\begin{center}
\includegraphics[width=12cm]{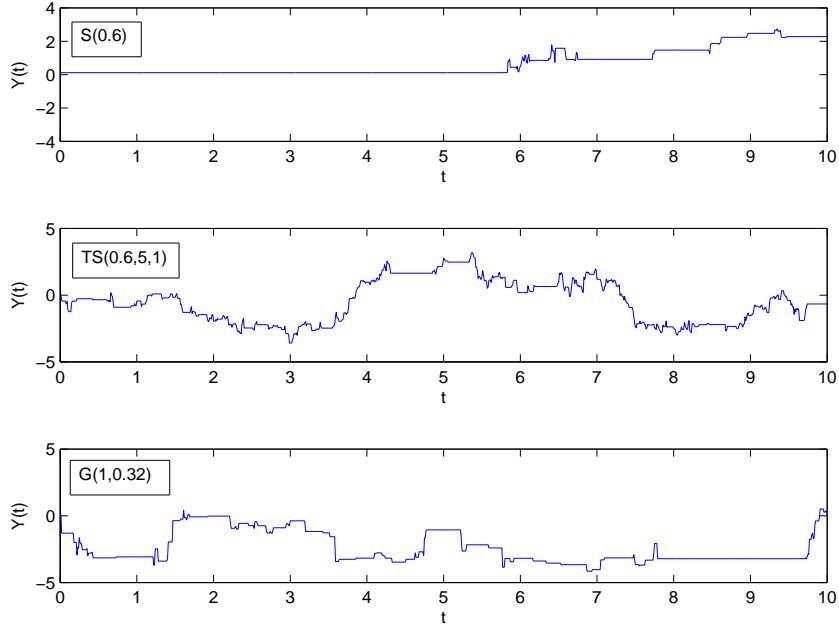}\label{traj}
 \caption{Sample trajectories of the subordinated process $\{Y(t)\}$ with the three considered subordinator distributions: S($\alpha$) -- $\alpha$-stable, TS($\alpha,\lambda,c$) -- tempered stable and G(a,c) -- gamma. The $\alpha$ parameters in the stable and tempered stable distributions are equal to $0.6$. The parameters of the gamma distribution are chosen so it's mean is equal to the mean of the tempered stable distribution.}
\end{center}
\end{figure}

Sample trajectories of the process $\{Y(t)\}$ defined in (\ref{sub1}) are plotted in Figure \ref{traj}. The chosen parameters correspond to the middle panels of Figure \ref{pdf}. 
Observe visible differences in the character of constant time periods.

Now, we focus on one of the most popular characteristic of the recorded process trajectories in experimental analysis, namely mean squared displacement. Recall that the ensemble averaged mean squared displacement is defined as:
\begin{equation}
<Y^2(t)>=\int_{-\infty}^{\infty}x^2P(x,t)dx,
\end{equation}
where $P(x,t)$ is the probability of finding a particle in a infinitesimal interval $(x,x+dx)$ at time $t$.
On the other hand, the time averaged mean squared displacement is given by:
\begin{equation}
\delta^2(t,T)=\frac{\int_0^{T-t} (Y(s+t) - Y(s))^2 ds}{T-t},
\end{equation}
where $T$ is the length of the analyzed time series.

For a standard Brownian motion the mean squared displacement (msd) scales as $t$ no matter if it is calculated as the ensemble or the time average. However, the behavior of the ensemble average changes under subordination scenario. In the $\alpha$-stable case ensemble averaged msd scales as $ t^{\alpha}$ \cite{metzler22}, while in the tempered stable case as $t^{\alpha}$ for $t\rightarrow 0$ and as $t$ for $t\rightarrow\infty$ \cite{woy,stanis}. It can be shown (for a detailed derivation see Appendix) that in the gamma case the ensemble average scales as
\begin{equation}
<Y^2(t)>\sim\left\{ \begin{array}{ccr}
-\frac{e^{-t/a}}{\log (t/a)} & \mbox{ as } & t\rightarrow 0,\\
t/a & \mbox{ as } & t\rightarrow \infty.
\end{array}
\right. 
\end{equation}

\begin{figure}[tb]
\begin{center}
\includegraphics[width=12cm]{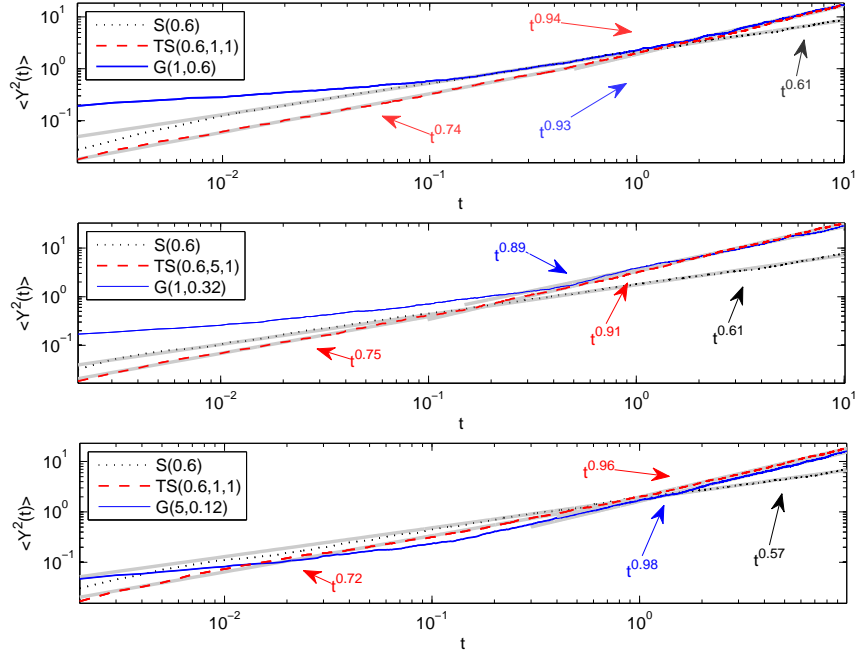}\label{msd}
 \caption{Mean squared displacement calculated as the ensemble average over 1000 trajectories of the subordinated process $\{Y(t)\}$ with the three considered subordinator distributions: S($\alpha$) -- $\alpha$-stable, TS($\alpha,\lambda,c$) -- tempered stable and G(a,c) -- gamma. The $\alpha$ parameters in the stable and tempered stable distributions are equal to $0.6$. The parameters of the gamma distribution are chosen so it's mean is equal to the mean of the tempered stable distribution. The fitted power law functions are plotted with the corresponding gray lines.}
\end{center}
\end{figure}

In Figure \ref{msd} we plot the mean squared displacement calculated as the ensemble average over 1000 simulated trajectories in the three considered cases. 
The chosen parameters are the same as on the corresponding panels of Figure \ref{pdf}. Moreover, we fit a power law function to each of the obtained curves, except for small $t$ in the gamma case. In the $\alpha$-stable case the power law is fitted for the whole range of $t$, while in the tempered stable case separately for small and large $t$ and in the gamma case only for large $t$. Observe that the obtained values are close to the theoretical power laws.

\begin{figure}[tb]
\begin{center}
\includegraphics[width=12cm]{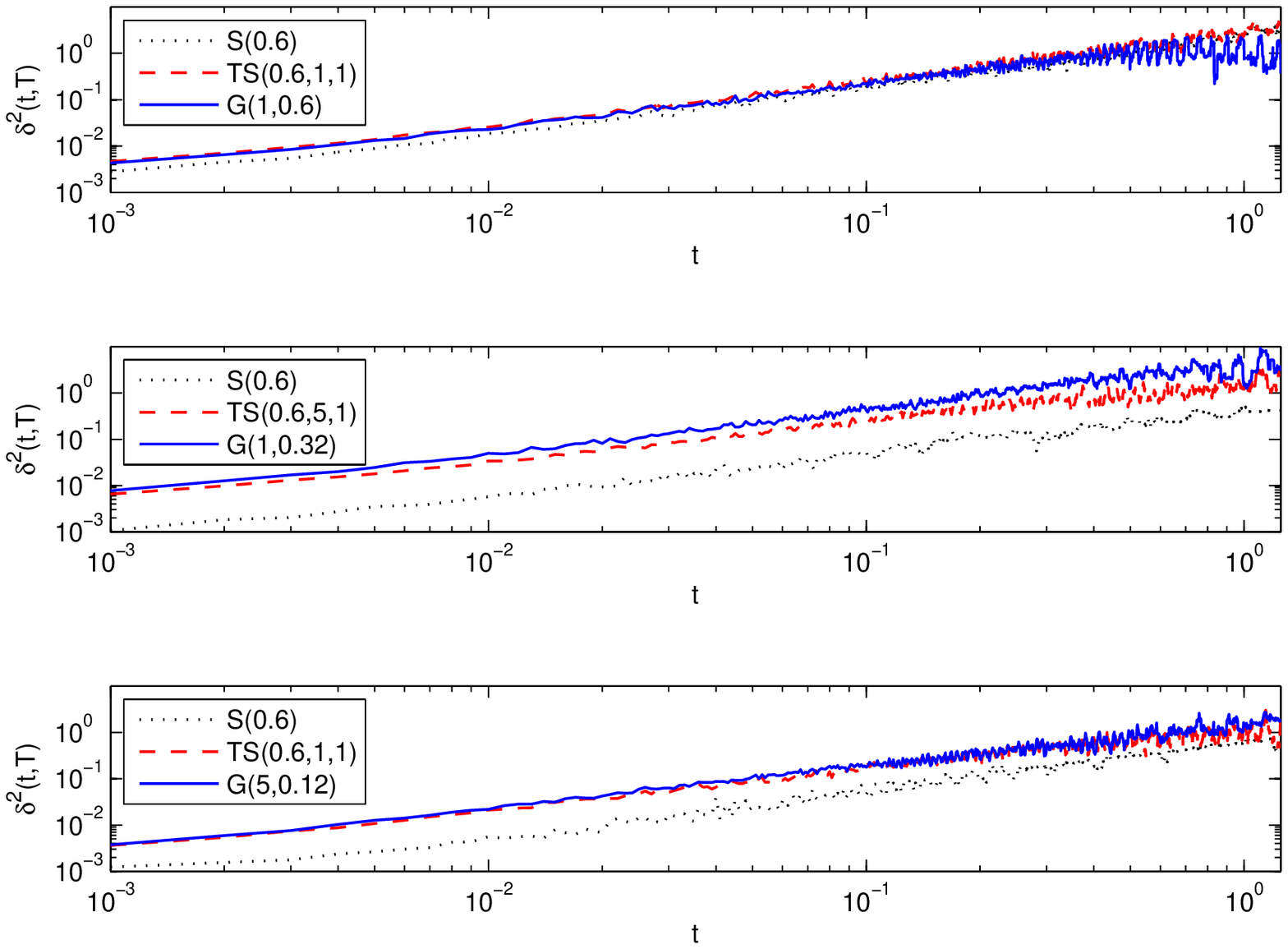}\label{msd_timeAvg}
 \caption{Mean squared displacement calculated as the time average of the subordinated process $\{Y(t)\}$ with the three considered subordinator distributions: S($\alpha$) -- $\alpha$-stable, TS($\alpha,\lambda,c$) -- tempered stable and G(a,c) -- gamma. The $\alpha$ parameters in the stable and tempered stable distributions are equal to $0.6$. The parameters of the gamma distribution are chosen so it's mean is equal to the mean of the tempered stable distribution.}
\end{center}
\end{figure}

Finally, we calculate the time averaged mean squared displacement. The obtained values calculated as the time average from a simulated trajectory of each of the three considered processes is plotted in Figure \ref{msd_timeAvg}. Observe that in all cases the obtained msd behaves as $ t$.

\section{Conclusions}
In this paper we have examined the anomalous diffusion models based on the subordinated Brownian motion with three types of subordinators distribution: $\alpha-$stable, tempered stable and gamma. The main result is related to the properties of the analyzed processes. We have pointed at the asymptotic behavior of the mean squared displacement in three considered cases and showed that in gamma case for small values of the arguments we obtain completely different behavior (non-power) from this observed in two other cases. 
\section*{Appendix}

In order to show the asymptotic behavior of the ensemble mean squared displacement (msd) in the gamma case we use the Proposition 3 in \cite{lageras}, namely if $\mu=<U(1)>$ is finite, then $<S(t)>\sim t/\mu$ for large $t$, where $\{U(\tau)\}$ and $\{S(t)\}$ are the subordinator and its inverse defined in (\ref{salpha}), respectively. In the gamma case with parameters $a$ and $c$, $<U(1)>=ac$ therefore when $t\rightarrow \infty$ we have
\[<Y^2(t)>=<B^2(S(t))>=<S(t)>\sim \frac{t}{ac}.\]
In order to show the asymptotic behavior of the msd function for small $t$ we use its  explicit form:
\begin{eqnarray}\label{msdgam}
<Y^2(t)>=\frac{1}{c}\int_{0}^{\infty}\frac{\gamma(\tau,t/a)}{\Gamma(\tau)}d\tau.
\end{eqnarray}
We can use the series expansion of the incomplete gamma function:
\begin{equation}
\gamma(s,x)=\Gamma(s)x^{1/2s}e^{-x}\sum_{n=0}^{\infty}e_n(-1)x^{1/2n}I_{n+s}(2x^{1/2}),
\end{equation}
where 
\begin{equation}
e_n(z)=\sum_{k=0}^{n}\frac{z^k}{k!}.
\end{equation}
and $I_v(z)$ is the modified Bessel function defined as follows:
\begin{equation}
I_v(z)=\left(\frac{1}{2}z\right)^v\sum_{k=0}^{\infty}\frac{\left(\frac{1}{4}z^2\right)^k}{k!\Gamma(v+k+1)}.
\end{equation}
The function $I_v(z)$ can be for small $z$ approximated by
\begin{equation}
I_v(z)\sim\frac{\left(\frac{1}{2}z\right)^v}{\Gamma(v+1)}.
\end{equation}
Therefore, when $t\rightarrow 0$, the function under the integral in (\ref{msdgam}) behaves like:
\begin{eqnarray}
\frac{\gamma(\tau,t/a)}{\Gamma(\tau)}&\sim& (t/a)^{1/2\tau}e^{-t/a}\sum_{n=0}^{\infty}e_n(-1)(t/a)^{1/2n}\frac{\left(t/a\right)^{1/2(n+\tau)}}{\Gamma(\tau+1+n)}\nonumber\\
&=&(t/a)^{\tau}e^{-t/a}\sum_{n=0}^{\infty}\frac{e_n(-1)(t/a)^{n}}{\Gamma(\tau+1+n)},\\
\end{eqnarray}
what gives
\begin{equation}
\frac{\gamma(\tau,t/a)}{\Gamma(\tau)}\sim\frac{(t/a)^{\tau}e^{-t/a}}{\Gamma(\tau+1)}.
\end{equation}
Thus, when $t\rightarrow 0$ we obtain 
\begin{eqnarray}\label{aproksymacja}
<Y^2(t)> \sim \frac{e^{-t/a}}{c} \int_{0}^{\infty}\frac{(t/a)^{\tau}}{\Gamma(\tau+1)}d\tau.
\end{eqnarray}
Now, note that the $\frac{1}{\Gamma(z)}$ function can be approximated as:
\begin{equation}
\frac{1}{\Gamma(z)}\sim\sum_{k=1}^{\infty}a_kz^k
\end{equation}
for some  $a_k$ that are independent of $z$ and satisfy the following relation:
\begin{equation}
a_n=na_1a_n-a_2a_{n-1}+\sum_{k=2}^n(-1)^k\zeta(k)a_{n-k},
\end{equation}
where $\zeta(z)$ is the Riemann zeta function \cite{riemann_zeta}.
As a consequence, we have:
\begin{equation}
<Y^2(t)> \sim \frac{e^{-t/a}}{c} \sum_{k=1}^{\infty}a_k\int_{0}^{\infty}(t/a)^{\tau}(\tau+1)^kd\tau.
\end{equation}
In order to simplify the notation denote $u=t/a$. We have
\begin{equation}
<Y^2(au)> \sim \frac{e^{-u}}{c} \sum_{k=1}^{\infty}a_k\int_{1}^{\infty}u^{\tau-1}\tau^kd\tau =\frac{e^{-u}}{cu}\sum_{k=1}^{\infty}a_k\int_{1}^{\infty}u^{\tau}\tau^kd\tau.
\end{equation}
Finally, let us consider the asymptotic behavior of the function $f(u,k)=\int_{1}^{\infty}u^{\tau}\tau^kd\tau$.
Integrating by parts gives the recursive relation:
\begin{equation}
f(u,k)=-\frac{u}{\log u}-\frac{k}{\log u}f(u,k-1),~~f(u,1)=-\frac{u}{\log u}\left(1+\frac{1}{\log u}\right).
\end{equation}
Therefore when $u\rightarrow 0$ the $f(u,k)\sim -\frac{u}{\log u}$, what yields the asymptotic behavior of the $<Y^2(au)>$ for small $u$, namely:
\begin{equation}\label{msdG}
<Y^2(au)> \sim -\frac{-e^{-u}}{c\log u} \sum_{k=1}^{\infty}a_k=A \frac{e^{-u}}{\log u},
\end{equation}
where $A=\mbox{const}$.
Substituting $u=t/a$ in (\ref{msdG}) we obtain:
\begin{equation}
<Y^2(t)> \sim -\frac{-e^{-t/a}}{c\log (t/a)} \sum_{k=1}^{\infty}a_k=A \frac{e^{-t/a}}{\log( t/a)}.
\end{equation}

\section*{Acknowledgments}
We are deeply grateful to Marcin Magdziarz for  stimulating
discussions  and his valuable suggestions. \\

The work of J.J was partially financed by the European Union within the European Social Fund.\\

\end{document}